\documentclass[twocolumn]{revtex4-1}
\usepackage{subfigure}
\usepackage{amssymb,amsthm,amscd, amsbsy, array}
 \usepackage{amsmath,bm,color}
\usepackage{graphicx}
\newcommand{\mathsym}[1]{{}}

\usepackage{epsfig}
\usepackage{epsf}
\usepackage{rotating}
\usepackage{color}

\newcommand{\half}{{\scriptstyle{\frac{1}{2}}}}

\newcommand{\lf}{\left (}
\newcommand{\lfq}{\left [}

\newcommand{\rg}{\right )}
\newcommand{\rgq}{\right ]}

\def\smallover#1/#2{\hbox{$\textstyle{#1\over#2}$}}
\def\beq{\begin{equation}}
\def\eeq{\end{equation}}
\def\beq{\begin{equation}}
\def\eeq{\end{equation}}
\def\bea{\begin{eqnarray}}
\def\eea{\end{eqnarray}}
\newcommand{\nn}{\nonumber}
\def\lf{\left(}
\def\rg{\right)}
\def\lq{\left[}
\def\rq{\right]}
\def\lgr{\left\{}
\def\rgr{\right\}}
\def\p{{\partial}}
\def\p{{\partial}}

\def\bn{{\bf n}}

\def\cE{{\mathcal E}}

\def\*{{\star}}

\def\freccia{{ \rightarrow }}

\def\aa{{{\`a }}}

\def\sign#1{{ \, \textrm{sign}\lq #1 \rq}}

\usepackage[authormarkup=none, markup=default, deletedmarkup=xout]{changes}

\definechangesauthor[color=purple]{3}
\definechangesauthor[color=blue]{vito}
\definechangesauthor[color=red]{1}
\begin{document}

 \title{Helicoids in Chiral Liquid Crystals under External Fields}
\author{G. De Matteis \dag\S \footnote{e-mail: giovanni.dematteis@istruzione.it},  $\quad$  L. Martina \dag \ddag \footnote{e-mail:martina@le.infn.it}, $\quad$ C.  Naya \ddag\footnote{e-mail: carlos.naya@le.infn.it}, $\quad$ V. Turco \dag \ddag \footnote{ e-mail: vito.turco@le.infn.it}
\\ \dag Dipartimento di Matematica e Fisica, Universit\aa del Salento\\  \ddag  INFN, Sezione di  Lecce, Via per Arnesano, C.P. 193 I-73100 Lecce, Italy\\ \S GNFM-INDAM, Citt\`a Universitaria - P.le Aldo Moro 5, C.P. 00185 Roma, Italy \\}
\date{\today}

%%%%%%%%%%%%%%%%%%%%%%%
\begin{abstract}
Cholesteric Liquid Crystals (CLCs), subject to externally applied magnetic fields
and confined between two parallel planar surfaces with strong homeotropic anchoring conditions,
are found to undergo transitions to different types of helicoidal configurations with disclinations.
Analytical and numerical studies are performed in order to characterise their properties. In particular,
we produce a phase diagram for the transitions from the nematic state to the helicoidal phases
in terms of the molecular chirality and the strength of the applied magnetic field.
\end{abstract}
  
  \maketitle
 
 \section{Introduction}

 In the absence of external applied magnetic or electric fields and in free space, a cholesteric liquid crystal
twists uniformly about a single axis: rod-like chiral molecules self-assemble into a helical arrangement along
a single direction.
Under confinement, the natural twisted structures of the chiral nematic liquid crystals are often incommensurate with the geometry, dimension and surface anchoring of the confining regions. Due to this geometric frustration, the confined chiral liquid crystals struggle to maintain their helical pitch and the preferred directors at the confining walls.
A similar frustration mechanism takes place for a cholesteric liquid crystal in an electric or magnetic field: the field alignment is incompatible with the cholesteric twist.
In both cases of frustration, the cholesteric is subject to an anisotropic environment: the competition
between favored twist and anisotropy leads to the distortion and the partial or total unwinding of the helical nematic texture and to the formation
of intriguing frustrated director configurations with topological defects \cite{1, 2}. The simultaneous interplay of the intrinsic molecular chirality, the external fields
and the boundary conditions, generates new structures, either localised or extended.
These structures can be elongated string-like objects called {\emph{cholesteric fingers (or threads)}} or {\emph{helicoids}}
\cite{4, baudry2, oswald1, oswald2, 3, 8},
or they may be localised objects called {\emph{cholesteric bubbles}} or {\emph{spherulitic domains}} \cite{tako, chandra, nawa, coche}.
Recently, it has been recognized that cholesteric bubbles can have the remarkable properties of {\emph{skyrmions}}.
Skyrmions, originally proposed in the field of nuclear physics \cite{skyrme},
are localised structures in which the magnitude of the order parameter (the nematic director $\mathbf{n}$ in this case)
remains constant, but the orientation continuously varies in a complex texture that cannot be annealed away.
Moreover, very recently, Smalyukh et al. generated in confined chiral liquid crystals a type
of defects in the director field configurations, called {\emph{triple-twist torons}}, by geometrical
frustration and by using Laguerre-Gaussian vortex laser beams.
They also showed numerically their existence from a theoretical point of view \cite{acke1, acke2, acke3}.

All such configurations, i.e. helicoids, skyrmions, torons, and other specific solitonic textures,
e. g. {\emph{hopfions}} and {\emph{merons}} \cite{acke1, acke2, acke3, pandey},
are stabilised by some type of topological and/or non-topological conservation laws \cite{13} and, at least in some approximate setting, they can
be described in terms of integrable nonlinear equations \cite{Spheruliti}.
Moreover, these novel structures have been studied through a range of techniques, including experiments
and numerical simulations in several confinement geometries such as thin layers, wedge cells, cylindrical cavities and spherical droplets \cite{lansac, lagemaat,fukudazumer,eun}.
Relevant variational and numerical calculations for skyrmions have been done in two dimensions by Bogdanov et al. \cite{bogdanov1, bogdanov2}, and more recently in three dimensions by Leonov et al. \cite{leonov2014}.
In the latter paper authors studied the Frank-Oseen free energy in the confined geometry
\begin{equation}
\label{regionB}
\mathcal{B}=\lbrace (x,y,z)\in\mathbb{R}^3,  \mid z\mid\leq \dfrac{L}{2}\rbrace,
\end{equation}
$L$ being the nematic cell gap, under the action of an external field and with weak anchoring conditions,
i.e. the Rapini-Popoular anchoring surface energy \cite{stewart_book}.
They calculated the director texture
for skyrmions. The boundary conditions translate into conditions on the partial derivatives of the director on the confining surfaces.
Later, in \cite{17}, the authors performed similar variational and numerical calculations in the same confining geometry but for the case
of strong anchoring ($\mathbf{n}$ perpendicular at the confining surfaces) and in the absence of external fields.
In particular the authors of \cite{17} worked out analytical and numerical solutions for a single helicoid and for
a helicoid lattice. By using similar calculations, they investigated isolated skyrmions
and skyrmion lattices.

In this paper, we consider a chiral liquid crystal confined within $\mathcal{B}$ as in (\ref{regionB})
with strong homeotropic anchoring conditions
and in the presence of an external magnetic field.

As opposed to \cite{17}, the inclusion of an external field leads to a nonlinear partial differential equation (PDE)
for the director orientation angle, more precisely the elliptic sine-Gordon equation on the strip, possibly with discontinuous boundary conditions.
New static chiral states are found and recognised to be similar to cholesteric fingers (helicoids)
with defects of disclination type \cite{4, 8}.  
We classify and describe these configurations by analytical and numerical methods.
In particular, we discuss a new type of solutions called $2\pi-$helicoids, where the director field $\mathbf{n}(\mathbf{r})$, independent of $y$,
rotates by $2\pi$ over the strip, that is, the projection of $\mathcal B$ onto the (x, z) plane. 
In addition to the $2\pi-$helicoids, we also find $\pi-$helicoids, where the director $\mathbf{n}(\mathbf{r})$ only twists once over the strip. All these solutions contain disclination-type singularities.
Accordingly, the evaluation of the static free energy leads to the introduction of a phenomenological cut-off, which,
in turn, determines a critical parameter for the transition to the nematic uniform state.
 Transitions from the uniform nematic state to $2\pi-$ and $\pi-$helicoids
are represented in the phase space of the spontaneous chiral twist and magnetic strengths.

The paper is organised as follows. In Sec. II, we introduce the mathematical model. In Sec. III and IV
we solve the nonlinear PDE problem and find the solutions in a closed analytical form.
In Sec. V, we build up a phase diagram of the solutions by energy comparison.
Finally, in Sec. VI, we draw our conclusions and plan future work. 

%%%%%%%%%%%%%%%%%%%%%%%%%%%%%%%%%%%%%%%%%%%%%%
%%%%%%%%%%%%%%%%%%%%%%%%%%%%%%%%%%%%%%%%%%%%%%
%%%%%%%%%%%%%%%%%%%%%%%%%%%%%%%%%%%%%%%%%%%%%%
 
\section{The model}

 We consider a static CLC layer, confined in between two identical planar surfaces placed at $z = \pm \frac{L}{2}$, and extending  to infinity in the orthogonal directions $\lf x, y \rg$ in a suitable   cartesian  reference system $\lf O, x, y, z \rg$, with orthonormal basis vectors $\mathbf{x},\mathbf{y},\mathbf{z}$.
 The system is described by the uni-modular director field $\mathbf{n}\lf \mathbf{r} \rg \in \mathbb{RP}^2$
\bea
\mathbf{n}\lf \mathbf{r} \rg =\lf \sin\theta(\mathbf{r})\cos\phi(\mathbf{r}),\sin\theta(\mathbf{r})\sin\phi(\mathbf{r}), \cos\theta(\mathbf{r})\rg ,
\eea
due to the $Z_2$ symmetry of the microscopic model \cite{deGennes}, and governed  by  the Frank-Oseen  free energy density
 \bea
&\cE_{FO} =&   \frac{K_1}{2}\left(\nabla\cdot\mathbf{n}\right)^2 + \frac{K_2}{2}\left(\mathbf{n}\cdot\nabla\times\mathbf{n}-q_0\right)^2 \nn \\ &&+ \frac{K_3}{2} \left(\mathbf{n}\times\nabla\times\mathbf{n}\right)^2  -\frac{\chi_a}{2}\lf \mathbf{n}\cdot \mathbf H \rg^2 ,   \label{FrankOseenEn}
\eea

 \noindent where $q_0$ is the spontaneous chirality constant of the cholesteric phase,  while the positive reals $K_i $ denote the splay, twist and bend Frank elastic constants, respectively,  for  which   we  use the simplifying  one constant  approximation
$  K = K_1 = K_2 = K_3. $ The last term  represents the interaction energy density with an external static magnetic field  $\mathbf H= \text{H} \; \mathbf{z}$, which  is assumed to be  uniform. 
  
 At  the bounding surfaces,  we impose strong homeotropic anchoring conditions, \emph{i.e.}
$\mathbf{n}\lf x, y, z = \pm \frac{L}{2}\rg = \mathbf{z}. $

Both $\mathbf{H}$  and  the confinement break  the general rotational and  translational symmetry along the $\mathbf{z}$ direction of the fundamental cholesteric helices, \emph{i.e.} $ {\bf  n}\lf {\bf r} \rg=\lf 0, \sin q_0x, \cos q_0x\rg$.  Hence, cholesteric helices are deformed,   possibly  leading to extended structures called  helicoids, or to localized cholesteric bubble domains, called spherulites, which have been considered in \cite{Spheruliti, PRE}.

The special symmetry reduction (constant $\phi=-\pi/2$ and $y$ invariance) \beq {\bf  n}\lf {\bf r} \rg = \lf 0, - \sin \theta\lf x,\, z\rg, \cos \theta\lf x,\, z\rg\rg, \eeq
for  the director field leads to  the cholesteric finger  phase
with its axis along the $ \mathbf{x}$ direction and   it simplifies the Frank-Oseen  energy (\ref{FrankOseenEn})   to
\bea E_{FO-2d}&=& \frac{K}{2}\int_{-\frac{L}{2}}^{\frac{L}{2}} dz \int_{-\infty}^{\infty} dx\; \lfq \left(\partial_x\theta (x,z)\right)^2 + (\partial_z\theta (x,z))^2\right.\nn\\ & &+ 2 q_0 \partial_x\theta (x,z)\left. + \frac{\chi_a \text{H}^2}{K} \sin^2 \theta(x, z)  \rgq, \label{FO-2d}\eea

\noindent where the constant magnetic contribution of the nematic phase has been subtracted.

The corresponding equilibrium equation is the elliptic sine-Gordon,
 \beq \p_{x}^2 \Theta + \p_{z}^2 \Theta = \Lambda^2 \sin \Theta, \quad  \Theta = 2 \theta  ,\quad \Lambda  = \sqrt{\dfrac{\chi_a}{K}} \text{H}, \label{SineGordon}\eeq

\noindent where $\Lambda$ is the reciprocal of the magnetic coherence length. Please note that this equation does not depend on $q_0$ as the term $K q_0 \partial_x \theta$ in (\ref{FO-2d}) is actually a null Lagrangian.
 
Large classes of solutions in the plane to equation (\ref{SineGordon}) are well known in the literature \cite{Borisov3} (and references therein),  but here we are dealing with different boundary conditions.   Our reduction is the most natural extension of the problem considered in \cite{17} and studied also in \cite{Lee92} in a linear setting. Here we study the nonlinear problem. Strong homeotropic anchoring conditions require $\Theta\lf x, \pm \frac{L}{2} \rg= 2 k \pi$ and $\Theta\lf x, \pm \frac{L}{2} \rg= 2 k' \pi$, with $k, \, k' \in \mathbb{Z}$, for negative and positive $x$ respectively. Thus, any non constant solution must have at least a jump singularity on the  boundaries.
 \begin{figure}[ht]\begin{center}\includegraphics[width=0.5\textwidth]{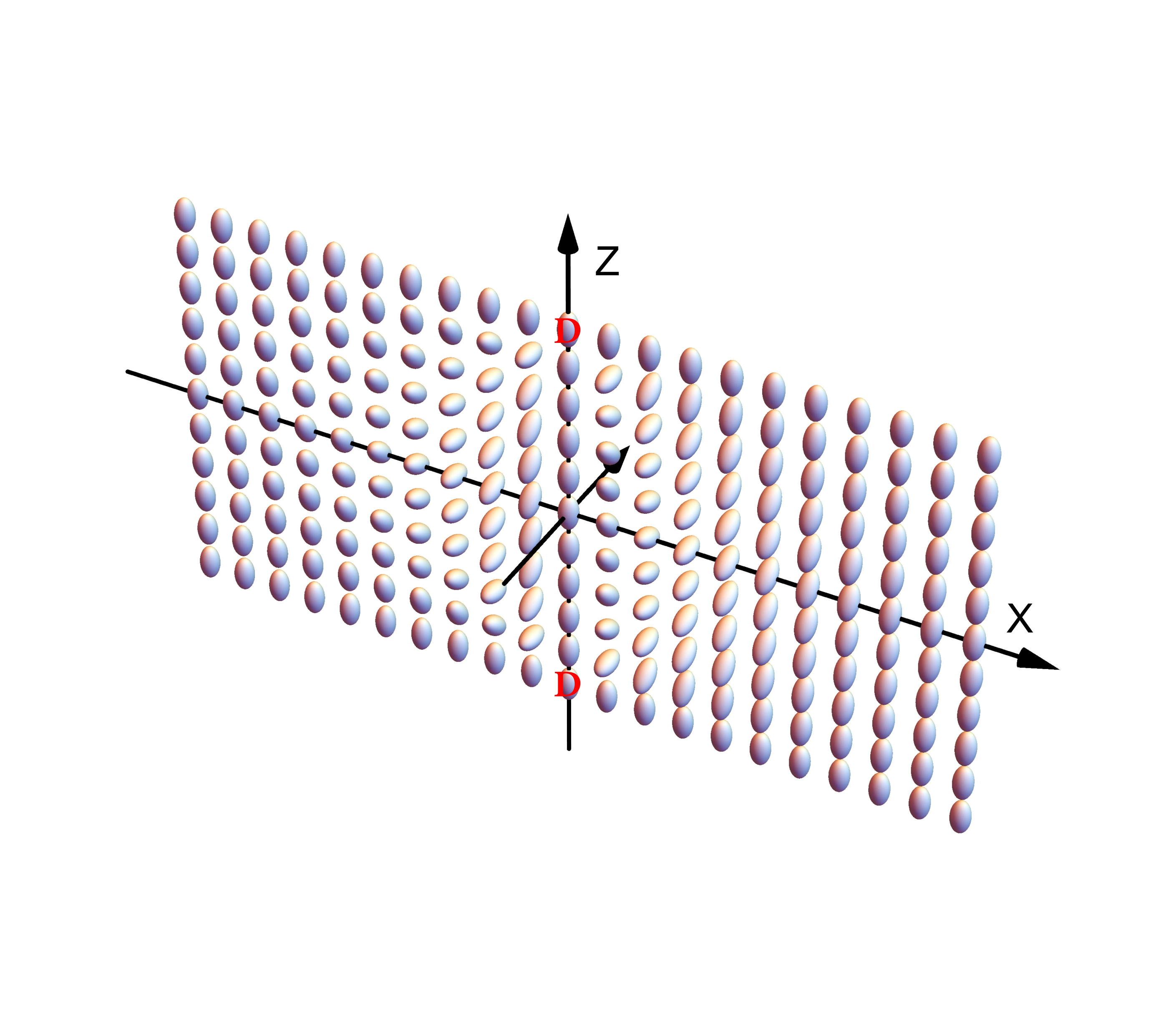}
\caption{ Distribution of  $\mathbf{n}\lf \mathbf{r} \rg$  for the  $2\pi$-helicoid, given in  (\ref{sol2pihelic}) for $n=0$. The picture shows a  cross section of the  configuration at $y = 0$ for $\Lambda L = 1$. The disclination is indicated by the letter D.} \label{2pihelicoidsFig}
\end{center}\end{figure}

%%%%%%%%%%%%%%%%%%%%%%%%%%%%%%%%%%%%%%%%%%%%%%
%%%%%%%%%%%%%%%%%%%%%%%%%%%%%%%%%%%%%%%%%%%%%%
%%%%%%%%%%%%%%%%%%%%%%%%%%%%%%%%%%%%%%%%%%%%%%

\section{$\boldsymbol{2 \pi}$ - helicoids}
In order to embody the above requirements, a known group-theoretical procedure  \cite{Kaptsov} suggests a simple ansatz  for the  solution, corresponding to the form  \beq  \Theta = 4 \arctan \lq X\lf x \rg Z\lf z \rg \rq, \label{arctanSol}\eeq 
which separately depends on  the $x$ and $z$ variables (see also \cite{Burylov}). 

Plugging this expression into (\ref{SineGordon}) and recalling the trigonometric identity $\sin(\arctan{x})=\frac{x}{\sqrt{1+x^2}}$, we get

\beq
\frac{X''}{X Z^2}+X^2\left(\frac{Z''}{Z}-\frac{Z'^2}{Z^2}\right)=X X'' -2 X'^2+\Lambda ^2 X^2+\frac{Z''}{Z^3}-\frac{\Lambda ^2}{Z^2},
\label{SG1}
\eeq%}}
where $(\cdot)^{'}$ and $(\cdot)^{''}$ indicate the first and the second derivatives with respect to the argument, respectively.

The integration procedure of this equation is well known \cite{Burylov}, but for self-consistency of the article and for convenience of the readers, we just sketch the main steps.

Then, performing the mixed derivative $\p^2_{xz}$ to both sides of eq. \eqref{SG1}, the \emph{r.h.s.} vanishes and it becomes
\beq
2X X'\lq\frac{Z''}{Z}-\frac{Z'^2}{Z^2}\rq^{'}-\frac{2  Z'}{Z^3}\lq\frac{X''}{X}\rq^{'}=0.
\eeq
Last equation can be separated in 
\beq
\frac{1}{X' X}\lq \frac{X^{''}}{X}\rq^{'} = 4a,\label{separation1}
\eeq
\beq
 \frac{Z^3}{Z^{'}}\lq \frac{Z^{''}}{Z}-2\lf \frac{Z^{'}}{Z}\rg^2\rq^{'}=4a,
\label{separation2}
\eeq
\noindent where $a$ is a separation constant.

Integrating eq. \eqref{separation1} and \eqref{separation2} with respect to $x$ and $z$, one obtains
\beq
X^{''}=2aX^3+b X, 
\eeq
\beq \label{separated}
 \frac{Z^{''}}{Z}-2\lf \frac{Z^{'}}{Z}\rg^2+\frac{2 a }{Z^2}+d=0,
\eeq
which can be further integrated (\emph{i.e.} multiplying the equation for $Z(z)$ in \eqref{separated} by ($Z'/Z^3$)) to get
\beq
(X^{'})^2=a X^4+b X^2 +c,
\label{separated2nd}
\eeq
\beq
 (Z^{'})^2=g Z^4+d Z^2 +a.
\label{separated1st}
\eeq
Equations \eqref{separated2nd} and \eqref{separated1st} contain five constants (the separation constant $a$ and the integration constants $b,c,d,g$), but only three of them are independent. Indeed, substituting them in eq. \eqref{SG1}, one can easily obtain the relations
\beq
g=c, \quad b=\Lambda^2-d.
\nn
\eeq
Thus, equations \eqref{separated2nd} and  \eqref{separated1st} can be written in the final form
\beq 
(X^{'})^2=a X^4+(\Lambda^2 -d) X^2 +c, \label{Xeq}
\eeq
\beq\label{Zeq}
 (Z^{'})^2=c Z^4+d Z^2 +a,
\eeq

\noindent which are solvable in terms of Jacobi elliptic functions.
 
 The step-like conditions on the boundaries force to have $Z\lf \pm \frac{L}{2} \rg = 0$ and to look for  functions $X\lf x \rg$  monotonic and unbounded, also at some finite point. 
    Then, one obtains 
   \beq X\lf x \rg = \pm  \sqrt{\frac{\Lambda^2-d}{ a}}\;\text{csch}\left( \sqrt{\Lambda^2-d }\, x \right) \label{cschsol}, \eeq
where $a > 0$ and  $\Lambda^2  >  d$.
   
   As for $Z(z)$, the compatibility of equation (\ref{Zeq}) with the sine-Gordon  (\ref{SineGordon}) and the above boundary conditions, set $d=-\frac{\pi^2}{L^2}$, in order to have real $\theta$ with the semi-period  of the corresponding $Z(z)$ exactly equal to the thickness $L$ of the sample. Thus, we are led to the general expression 
  \bea \theta_n &=& 2\,  \arctan\left[\frac{c_n\, \ell   }{\pi \lf 1+2 n \rg }\frac{\cos
   \left(\frac{\pi \lf 1+2 n \rg  
   z}{L}\right)}{
   \sinh\left(c_n \, \ell \, \frac{x}{L} \right)}\right] - \pi \sign{x}, \nn \\ \label{sol2pihelic}\eea 
  for $ n \in \mathbb{N}$, and where the effective scale $\ell = \Lambda \, L $  and  the modal factor $c_n = \lfq 
   1 +\frac{\lf 1+2 n \rg^2\pi
   ^2}{\ell^2} \rgq^{\half}$ have been introduced. The asymptotic behaviour of this family of solutions is  $\theta \stackrel{x \to \pm \infty}{\longrightarrow}{} \mp \pi$, that is, the director field $\mathbf{n}(\mathbf{r})$ rotates by $2\pi$ over the strip, as depicted in Fig. \ref{2pihelicoidsFig}.

We first observe that the solution only depends on the structural parameters $L$ and $\Lambda$ via their product $\ell$. Second, there exists an entire spectrum of excitations indexed by  $n \geq 0$.   Since for $n > 0$ solutions are discontinuous at $x=0$ and $\cos(\pi (1+2n) z/L) \leq 0$, the  most physically meaningful solution corresponds to $\theta_0\lf x, z \rg$ ( Fig. \ref{2pihelicoidsFig}), which is continuous and differentiable  at all points of the strip but at the special points $\lf x = 0, z = \pm \frac{L}{2} \rg$. Here it exhibits a discontinuity of $2 \pi$, indicating the presence of a disclination along the $\mathbf{y}$ direction.
 
Whilst the dependency on $z$ is determined uniquely by $L$, on the $x$ variable the typical scale  is $ \Lambda^{-1}\lf{1 + \frac{ \pi^2}{\ell^2}}\rg^{-\half}$,  
which comes from the tendency of the external magnetic field to align the molecules along its direction, further  enhanced  by the anchoring of the bounding surfaces.  

We also notice that solution \eqref{sol2pihelic} can be obtained by applying the nonlinear Bianchi superposition formula \cite{Bianchi} for
two 1-kink solutions of the elliptic sine-Gordon (\ref{SineGordon}). Such a theorem states that if $\Theta$, $\Theta_1$ and $\Theta_2$ are three solutions related by the overdetermined first order system, the so-called B\"acklund transformation, 
\bea
(\partial_x - \imath \partial_y) \lf \frac{\Theta_i-\Theta}{2} \rg &=& \beta_i \Lambda\; \sin\lf \frac{\Theta_i+\Theta}{2} \label{BT1} \rg \\
(\partial_x + \imath \partial_y) \lf \frac{\Theta_i+\Theta}{2} \rg &=& \frac{\Lambda}{\beta_i}\; \sin\lf \frac{\Theta_i-\Theta}{2} \rg ,\label{BT2}
\eea
($\beta_i \in \mathbb{C}$, $i=1, 2$),  then a  fourth solution is given by 
\beq \bar \Theta = \Theta + 4 \arctan \lfq \frac{\beta_2 + \beta_1}{\beta_2 - \beta_1} \tan \lf \frac{\Theta_2 -\Theta_1}{4}\rg\rgq. \label{superposForm} \eeq
As an application, one solves the system  \eqref{BT1}, \eqref{BT2} corresponding to the trivial solution   $\Theta =0$. This leads to the kink solutions $\Theta_i = 4 \arctan\left(e^{a_i+\frac{\beta_i \Lambda \zeta }{2}+\frac{\Lambda  \bar{\zeta}}{2 \beta_i }}\right)$, with  $a_i$ integration constants for $i =  1, 2$. Combining these two expressions into formula \eqref{superposForm} and requiring real solutions for real variables, one is led to the constraints $\beta_2 = \beta_1^{-1}$ and $a_1 = 0, \; a_2 = \imath \pi $. Furthermore, imposing the boundary condition one arrives at the expression (\ref{sol2pihelic}) with $\beta_1 = c_n + \sqrt{c_n^2 -1} $.

 \begin{figure}[t]
\centering
\includegraphics[width=.5\textwidth]{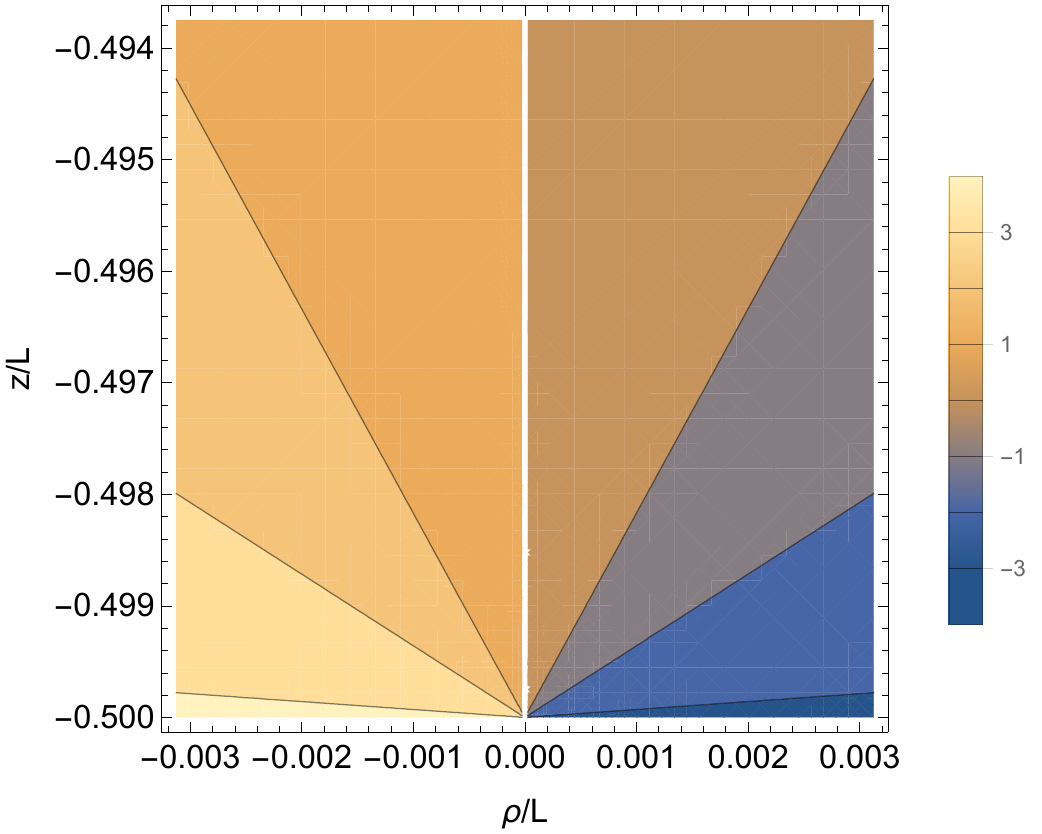}
\caption{Contour plot of $\theta_0$ around the disclination located at $x=0$, $z=-L/2$.}
\label{fig:disclination}
\end{figure}

This result suggests that multiple $2\pi$-helicoid solutions may be built up by iterating the process. Unfortunately, the 
boundary conditions we impose are too rigid and they lead to singular solutions, which will be excluded from further discussions.  

Finally, let us observe that since the total change of  $\theta$ is $2\pi$, we can distinguish two regions where the director rapidly passes through $\frac{\pi}{2}$, sandwiched by three different regions in which the liquid crystal is close to the uniform state.  Around the disclination,  $\theta_0$ has a conformally invariant behaviour which,  in cylindrical coordinates  $ x = \rho \cos t,\; z = -\frac{L}{2} + \rho \sin t  $,  can be expressed as a series with respect to $\rho$, namely,
   \bea  \theta_0 = -\sign{\cos t}\pi &+& 2 t-\frac{ \sin (2 t) \left(  \ell^2 \cos^2 \, t+ \pi ^2\right)}{6} \lf \frac{\rho }{L}\rg^2  \nn \\  & &+ O\lf\lf\frac{ \rho}{L}\rg^4\rg , \qquad  0\leq t \leq \pi .  \label{Cylexpansion}\eea
 Thus, $\theta_0$ is almost independent of the  distance $\rho$  from the disclination as shown by the contour plot in Fig. \ref{fig:disclination}. The  corresponding singularity in the energy  signals  a loss of order, and therefore, physically,  in the vicinity of the disclination, the  liquid crystal is melted from the ordered  to the disordered isotropic phase. To avoid the singularity, we single out a small semi-disk, of radius $a  \ll \frac{L}{2}$, around the disclination and  replace it with a region in the isotropic phase, where  it is common practice to uniformly impose an  energy density cut-off $ \cE_{max} \approx \frac{K}{ a^2 }$  with $a\approx 10^{-2} - 10^{-3} L  $ \mbox{\cite{Lee92}}. 
 
By using analytical solution \eqref{sol2pihelic}, we can write down an expression for the difference of energy with respect to the nematic phase in the limit of small $a/L$, which, up to second order, reads
\begin{widetext}
\begin{eqnarray}
&\Delta E_{2\pi} = E_{2\pi} - E_{\rm nem} \approx &\frac{K}{3} \bigg[ 24 \, G + 10 \, a^2 \Lambda^2 + 12 \, \sqrt{\pi^2 + \Lambda^2 L^2} + 3 \, a \, q_0 \log(16) - 2 \pi \bigg(6 -3 \, a \, q_0 + 3 L q_0 + 2 \, a^2 \Lambda^2 \nonumber   \\
&& + 6 \log \bigg(\frac{a}{2 L} \bigg) + 3 \log \Big( \Lambda^2 L^2 + 2 \pi ( \pi + \sqrt{\pi^2 + \Lambda^2 L^2}) \Big) \bigg) \bigg],\label{DeltaE}
\end{eqnarray}
\end{widetext}
\noindent where $G$ is the Catalan constant. Please note that $E_{2\pi}$ corresponds to the energy of the $2\pi$-helicoids while $E_{\rm nem}$ stands for the energy of the nematic phase, which equals zero as can be seen from eq. (\ref{FO-2d}).

At thermal equilibrium, stable 2$\pi$-helicoids  may exist  if  $ \Delta E_{2\pi} <0$; from the latter condition, one can infer a critical transition value for the chiral twist parameter as a function of the external field $\Lambda$. Notice that due to the presence of the external field,   expansion \eqref{DeltaE} for small $a/L$ holds as far as $\Lambda L$ is not very high, which  is true in the range of interest.  By neglecting the $(a/L)^2$ contributions in \eqref{DeltaE} and introducing the cut-off energy, we arrive at 
\begin{widetext}
\beq
\Delta E_{2\pi} = 2 K L (q_H - q_0) \left(\pi - \frac{1}{L} \sqrt \frac{K}{\cE_{max}} \left(\pi + \log (4) \right) \right),
\eeq
\noindent with

\begin{eqnarray}
q_H &= & \frac{4 \, G - 2 \pi + 2 \sqrt{\pi^2+\Lambda^2 L^2} + \pi \log \left(4 L^2 \, \frac{\cE_{max}}{K} \right) - \pi \log \left(\Lambda^2 L^2 + 2 \pi (\pi + \sqrt{\pi^2 + \Lambda^2 L^2})\right)}{L \left(\pi - \frac{1}{L} \sqrt \frac{K}{\cE_{max}} \left(\pi + \log (4) \right) \right)} ,
\end{eqnarray}
\end{widetext}
\noindent the critical chiral twist.
Since  the disclination radius has an important effect in locating the phase transition, $q_H$   might be used  to estimate  the effective size of the defects  appearing in  real liquid crystal samples.

%%%%%%%%%%%%%%%%%%%%%%%%%%%%%%%%%%%%%%%%%%%%%%
%%%%%%%%%%%%%%%%%%%%%%%%%%%%%%%%%%%%%%%%%%%%%%
%%%%%%%%%%%%%%%%%%%%%%%%%%%%%%%%%%%%%%%%%%%%%%
   
\section{$\boldsymbol{\pi}$ - helicoids} \label{PiHelicoids}
%%%%%%%%%%%%%%%%%%%%%%%

In a  ``more elementary''  class of  helicoids  the director $\bn \boldsymbol{(r)}$ twists only once, when  going from the vacuum state $\theta = \pi$ at $x \to - \infty$,  to the nearest $\mathbb{Z}_2$ equivalent  one, namely $\theta = 0$ for  $x \to + \infty$.   Moreover, because of the homeotropic anchoring, the boundary conditions on the confining surfaces will be 
\beq \theta\lf x, z = \pm \frac{L}{2} \rg = \lgr 
\begin{array}{cc}
 \pi  &  x<0     \\
0  &   x>0     
\end{array}
\right. .
\label{pibvpSineGordon}\eeq
Then,  in such a kind of solutions, which we  call $\pi$-helicoids,  a couple of disclinations appear on the boundaries, parallel to the $y$ axis.
  \begin{figure}[b]\begin{center}\includegraphics[width=0.5\textwidth]{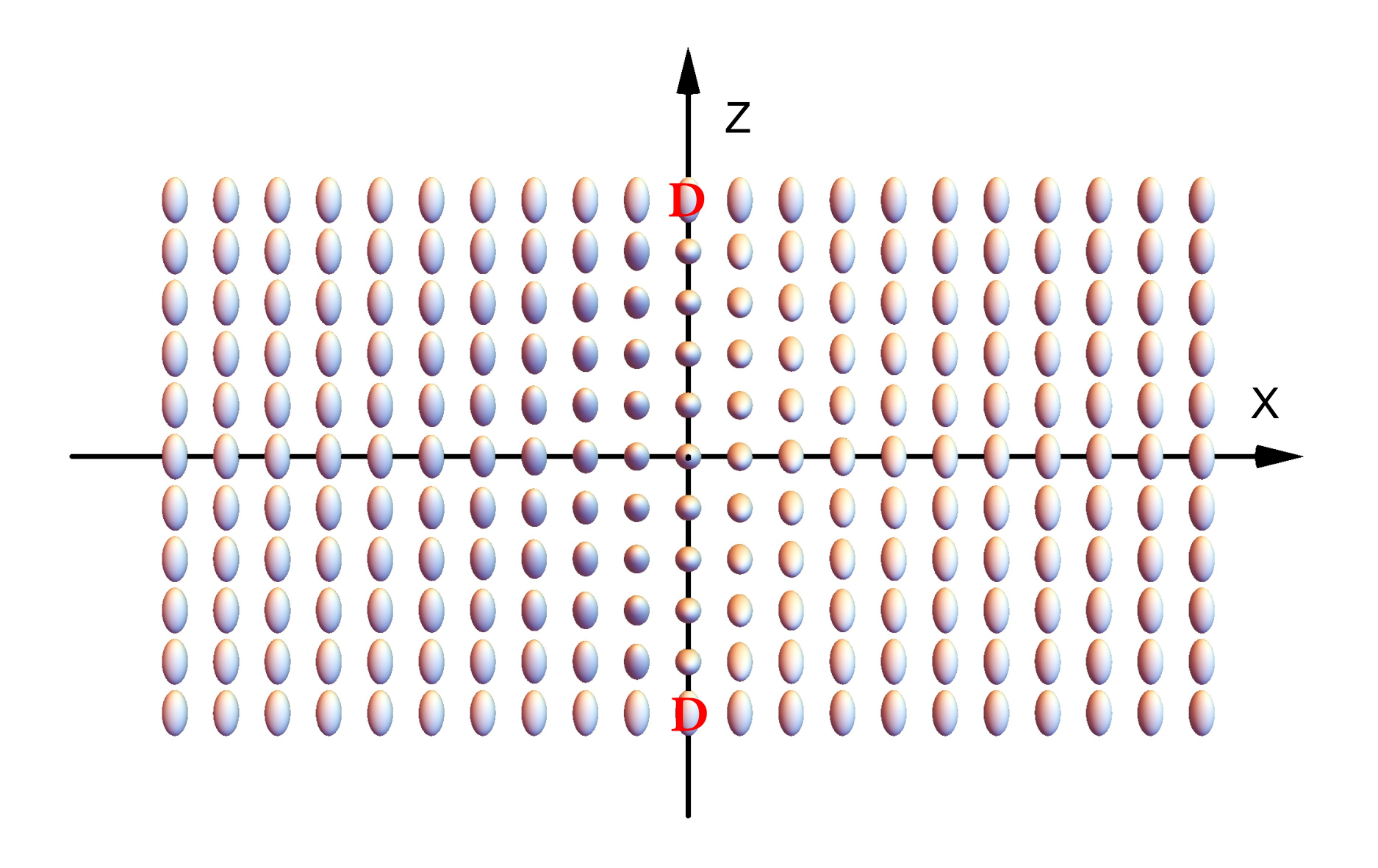}
\caption{ Distribution of the $\mathbf{n}\lf x, z \rg$ field for the  $\pi$-helicoid. The picture shows a  cross section of the  configuration at $y = 0$ for $\Lambda L = 1$. The disclination is indicated by the letter D.} \label{piHelicoidsFig}
\end{center}\end{figure} 
The analysis of the full nonlinear  problem \eqref{SineGordon} with  boundary values  (\ref{pibvpSineGordon}) is quite involved, even if one would  profit from the integrability properties of the sine-Gordon equation through its Lax pair formulation, as  developed in \cite{AKNS} and references therein (see Appendix for more details). Actually, the Lax pair problem is a matrix  linear representation of the overdetemined  B\"acklund system  \eqref{BT1} and \eqref{BT2}, and it is equivalent to finding a matrix integration factor $\Psi \lf x, z; \lambda \rg$,  depending on an auxiliary complex parameter $\lambda$, which makes the following 1-form exact
\bea \label{W_form} W &=& e^{  \Omega\lf \lambda \rg x +  \omega\lf \lambda \rg z \frac{{\hat \sigma_3}}{2} }\lf Q\lf x, z, \lambda\rg \Psi \, dx +  \imath Q\lf x, z, -\lambda\rg \Psi\, dz \rg \nn \\
&= &d \lf \exp\lfq  \Omega\lf \lambda \rg x +  \omega\lf \lambda \rg z \frac{{\hat \sigma_3}}{2}  \rgq  \Psi\rg , \eea
with ${\hat \sigma_3} \cdot = \lfq \sigma_3, \cdot \rgq$,  $\sigma_i$ being the Pauli matrices, $\Omega(\lambda)$ and $\omega(\lambda)$ the dispersion functions on $\mathbb{C}_\lambda/\{0\}$ defined as
 \beq \label{omegas} \Omega\lf \lambda \rg = \frac{\imath \Lambda}{2}\lf \frac{1}{\lambda} - \lambda\rg, \qquad \omega\lf \lambda \rg = \frac{\Lambda}{2}\lf \frac{1}{\lambda} + \lambda\rg , \eeq
and $Q(x,z,\lambda)$ the \emph{potential} matrix 
\bea  Q\lf x, z, \lambda\rg  = &\frac{\imath}{4} 
&\left(
 \frac{\Lambda}{\lambda}\lf  1 -\cos \Theta \rg\, \sigma_3 \right. \nn\\
& +& \left.  \lf \Theta_x - \imath \Theta_z \rg \, \sigma_1 -   \frac{\Lambda}{\lambda} \sin \Theta \, \sigma_2 \right).
\eea
Then, one can check that the closure condition of $W$ implies that $\Theta$ has to satisfy the sine-Gordon equation  \eqref{SineGordon}. However, this type of construction also requires the knowledge of the derivatives normal to the boundaries, possibly containing singularities not included in (\ref{pibvpSineGordon}). These difficulties are not alleviated in the limit of the linear approximation, which can be obtained by letting $\Theta = 2 \theta \; \approx\; 0$ and $\Psi \; \approx \; \mathbf{ 1}_2$, namely
\bea \label{W_lin}  W_{\rm lin}& =& 2 e^{-  \Omega\lf \lambda \rg x -  \omega\lf \lambda \rg z } \lf \imath  \lf \theta_x - \imath \theta_z - \frac{\Lambda \theta}{\lambda}\rg \right.  \, dx \nn \\ &&  - \left. \lf \theta_x - \imath \theta_z  + \frac{\Lambda \theta}{\lambda}\rg \, dy \rg.\eea 

 To avoid  such an  addition of extra information about the derivatives, a long procedure was introduced in  \cite{FLP}.  Such a work led to a unifying approach in solving both linear and nonlinear boundary value problems, via a suitable integral mapping of the boundary data into the solution \cite{Unified}.

The closure condition of $W_{\rm lin}$ yields the modified Helmholtz equation which can be solved by different types of transform methods, thus bypassing the above mentioned problem of the derivatives on the boundary. By exploiting  
the  discrete mirror symmetry along the x-axes of our problem  \eqref{SineGordon} and \eqref{pibvpSineGordon}, the  modified Helmholtz boundary value problem on the semi-strip arises
\begin{subequations}
\label{eq:bvpthetap}
\bea  \p_{x}^2 \theta_+ + \p_{z}^2 \theta_+ &=&  \Lambda^2 \theta_+\;  \;\forall \;  x>0 , \label{bvpthetap:1} \\ \theta_+\lf x, z =  \frac{\pm L}{2} \rg = 0 &,&      \theta_+\lf x = 0^+, z \rg = \frac{\pi}{2}  \;  \;\forall \;  | z | <  \frac{L}{2} ,\nn\eea
\beq    \theta_- \lf x, z \rg =  \pi -     \theta_+ \lf - x, z \rg ,  \;\forall \;  x < 0 \; \textrm{and}  \; | z | <  \frac{L}{2}. \label{mirror} \eeq  
\end{subequations}

In principle, this system could be solved by standard Fourier series methods \cite{Morse}. Because of the linearity of the equation and the separability of  the independent variables, one can look for solutions of \eqref{eq:bvpthetap}  in the form $\theta_+ \lf x, z \rg = \sum_{k=0}^\infty X_k \lf x \rg Z_k \lf z \rg $, where $Z_k$ belong to an  orthonormal  basis of the $L^2_{\lfq - \frac{L}{2}, + \frac{L}{2}\right]}$  space. Taking into account  the vanishing of the series at $| z | = \frac{L}{2}$,
one sets  $Z_k =  \sqrt{\frac{2}{L}}\cos\left(\frac{\pi \lf 2 k +1\rg z}{L}\right)$. Assuming that the series is absolutely and uniformly convergent together with its derivatives,  its substitution into \eqref{eq:bvpthetap} can be computed by exchanging the sum with the laplacian operator, which will act term by term. Using the ortho-normalization of the $Z_k $,  for the generic coefficient one gets the equation
\beq X_k^{''} = \lf \Lambda^2 + \left(\frac{\pi \lf 2 k +1\rg }{L}\right)^2 \rg X_k. \eeq
The corresponding solutions are real exponentials, from which we only keep those vanishing as $x$ tends to infinity, namely $X_k = \xi_k \exp \lfq -\frac{x \sqrt{\pi
   ^2 (2 k+1)^2+\Lambda^2\, L^2}}{L} \rgq$. The amplitudes $\xi_k$ are simply determined by letting $x \freccia 0$, and using again the ortho-normalization of the $Z_k$ for the series $\sum_{k = 0}^\infty \xi_k Z_k = \frac{\pi}{2}$, we are led to the solution   
   \beq \theta_+ \lf x, z \rg  = 2 \sum _{k=0}^{+ \infty}
   \frac{\lf -1 \rg^k}{2 k+1} e^{-\frac{x \sqrt{\pi
   ^2 (2 k+1)^2+\ell^2}}{L}} \cos
   \left(\frac{\pi  (2 k+1) z}
   {L}\right) . \label{seriessol}\eeq This latter cannot be cast in a factorized form as  in \eqref{sol2pihelic}, unless when $\Lambda =0$. In this case, the series can be summed up to the closed form 
   \begin{equation}
   \theta_+(x,z) = \arctan \left[\frac{\cos\left(\frac{\pi z}{L}\right)}{\sinh \left(\frac{\pi x}{L} \right)} \right],
   \end{equation}
  \noindent which, upon using the trigonometric identity $\arctan u + \arctan v = \arctan \left( \frac{u + v}{1 - uv}\right)$, matches Eq. (10) in Ref. \cite{17}  up to a global sign due to the equivalence $\theta \rightarrow \pi - \theta$ in order to comply with their far-field values.
   
Using expression (\ref{mirror}) one can continuously complete the solution also for $x <  0$, except at the points $\lf x=0, z = \pm \frac{L}{2} \rg$.
 The exponential decaying in $x$  has the  characteristic length   $\Delta x \approx  \frac{2 L }{\sqrt{\ell ^2+\pi^2}}$.  
 
 However, close to $x = 0$  the linear approximation to the sine-Gordon might be unsatisfactory, first because $\sin \frac{\pi}{2} \neq \frac{\pi}{2}$, and second because there the series \eqref{seriessol} is not uniformly and absolutely convergent, implying all well known convergence problems   at the discontinuity points.  Moreover, the partial derivatives $\p_x \theta \lf x , z \rg  $ and $\p_z \theta \lf x , z \rg $ are diverging for $x \to  0^\pm$ and $z \to \pm \frac{L}{2}$. So the differentiability of $\theta$ on the segment  $\{x = 0, | z | < L/2\}$ is lost.

By finding  an integration factor associated to the 1-form $W_{\rm lin}$ in eq. (\ref{W_lin}), we can also adopt a different representation of the same solution (see Appendix for details)  as follows \cite{AF}
\begin{widetext}
\bea \theta_+(x,z)&=& \frac{-1}{2 \pi} \lq  \int_{-\infty}^0 e^{\Omega\lf \lambda \rg x + \omega\lf \lambda \rg \lf z+\frac{L}{2} \rg} G_1\lf \lambda \rg \frac{d\lambda}{\lambda} + \int_{\infty}^0 e^{\Omega\lf \lambda \rg x + \omega\lf \lambda \rg \lf z-\frac{L}{2} \rg} G_1\lf \lambda \rg \frac{d\lambda}{\lambda}\right.\nn \\ && \left. +
\int^{- \imath \infty}_0 e^{\Omega\lf \lambda \rg x + \omega\lf \lambda \rg \lf z+\frac{L}{2} \rg} G_2\lf \lambda \rg \frac{d\lambda}{\lambda}\rgq, \label{integralAFrepr}\eea
\end{widetext}
     where
 \beq G_1\lf \lambda \rg = \frac{\imath \pi}{4} \frac{1- \lambda^2}{1+ \lambda^2} \frac{e^{\omega\lf \lambda \rg L} - 1}{e^{\omega\lf \lambda \rg L} + 1}, \eeq
 \bigskip
 \beq \;G_2\lf \lambda \rg =  \frac{\imath \pi}{4} \frac{1- \lambda^2}{1+ \lambda^2} \lf 1 - e^{- \omega\lf \lambda \rg L} \rg , \eeq
  encode the information about the  boundary conditions. In particular, 
 the function $ G_1\lf \lambda \rg$ has the poles ( $\ell = \Lambda L$)
\beq P_G = \lgr - \imath \lfq \sqrt{\ell^2+(2  n+1 )^2 \pi^2}-\pi  (2 n+1)\rgq/L \rgr_{n \in \mathbb{Z}}. \eeq As expected, the corresponding residues lead to expansion \eqref{seriessol}, this bringing both approaches together. On the other hand, by looking at the analytic properties of the integrands in the lower half-plane $\mathbb{C}_\lambda$ , one may suitably deform the integration contour  in formula  \eqref{integralAFrepr}, in order to extract more information about the solution near the 
\onecolumngrid
\begin{center}
\begin{figure}[t]
\subfigure[]{\includegraphics[width=.45\textwidth]{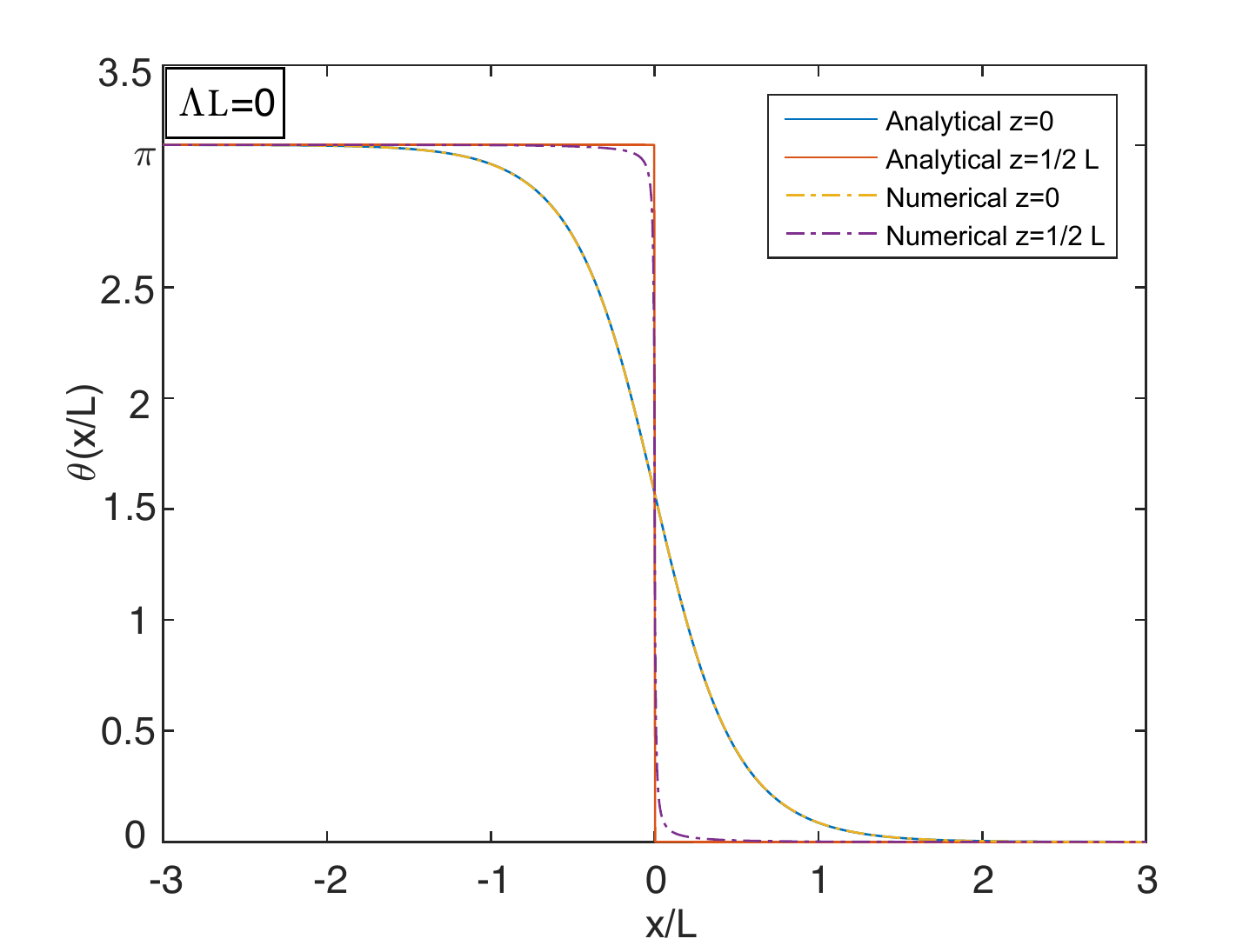}}
\subfigure[]{\includegraphics[width=.45\textwidth]{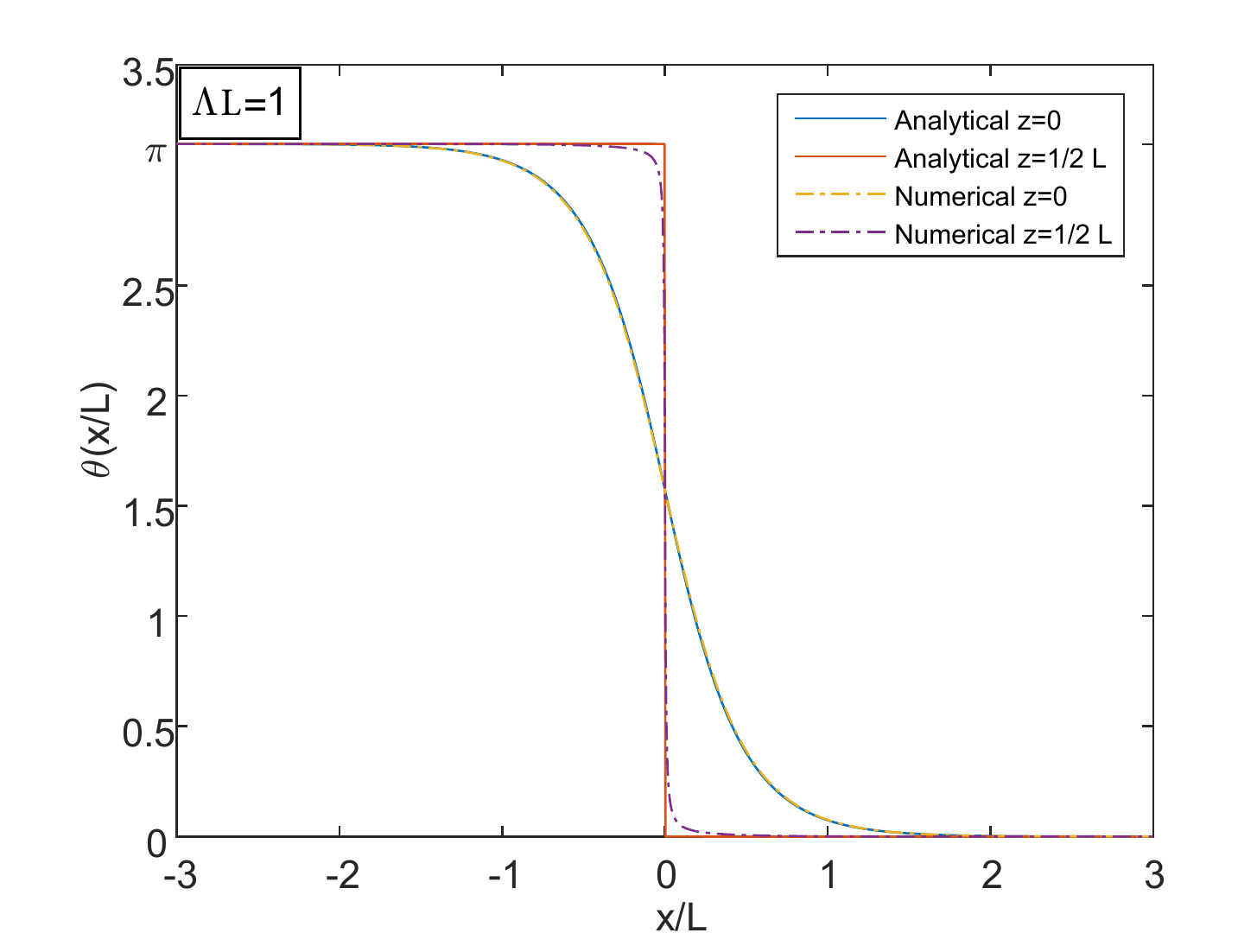}}
\subfigure[]{\includegraphics[width=.45\textwidth]{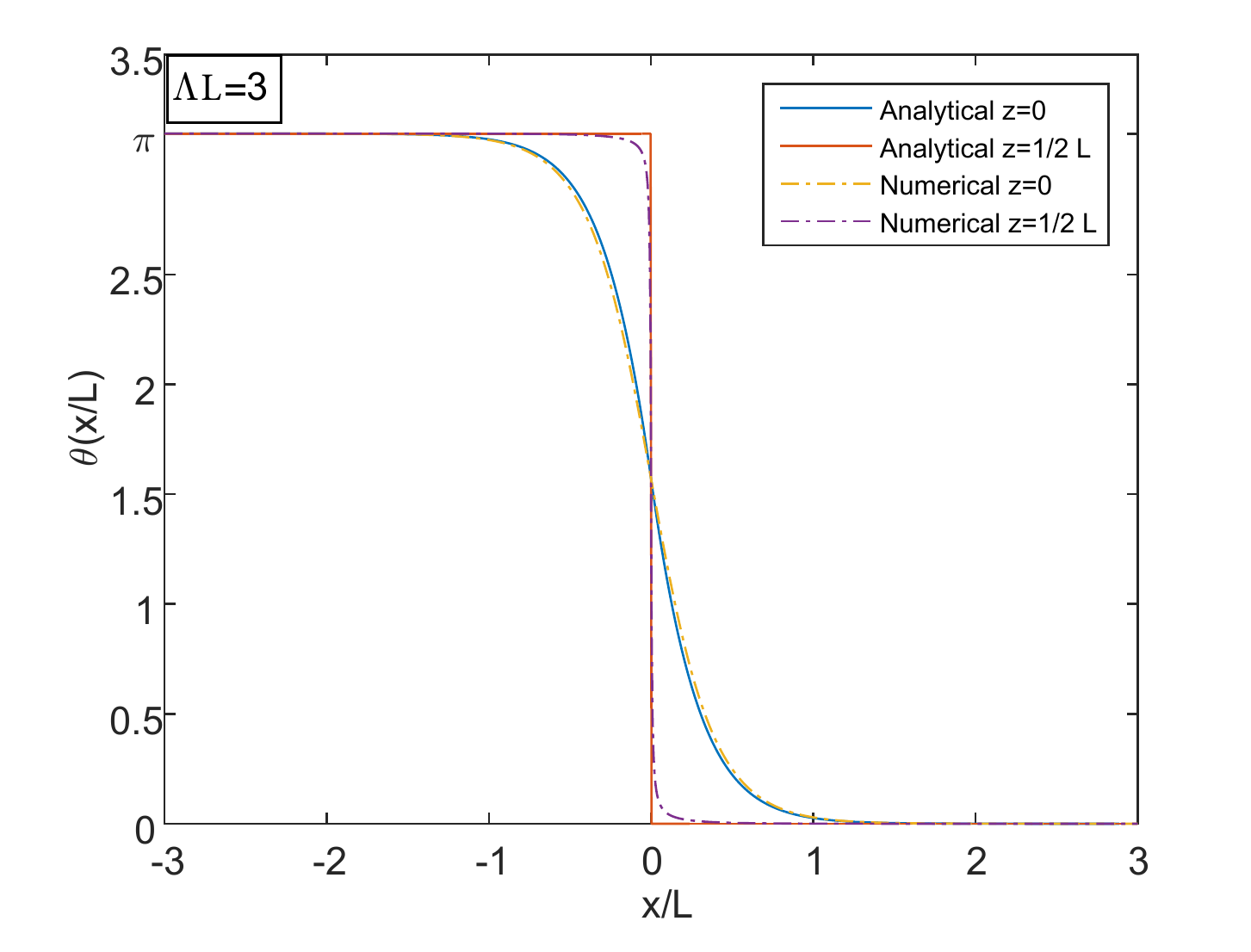}}
\subfigure[]{\includegraphics[width=.45\textwidth]{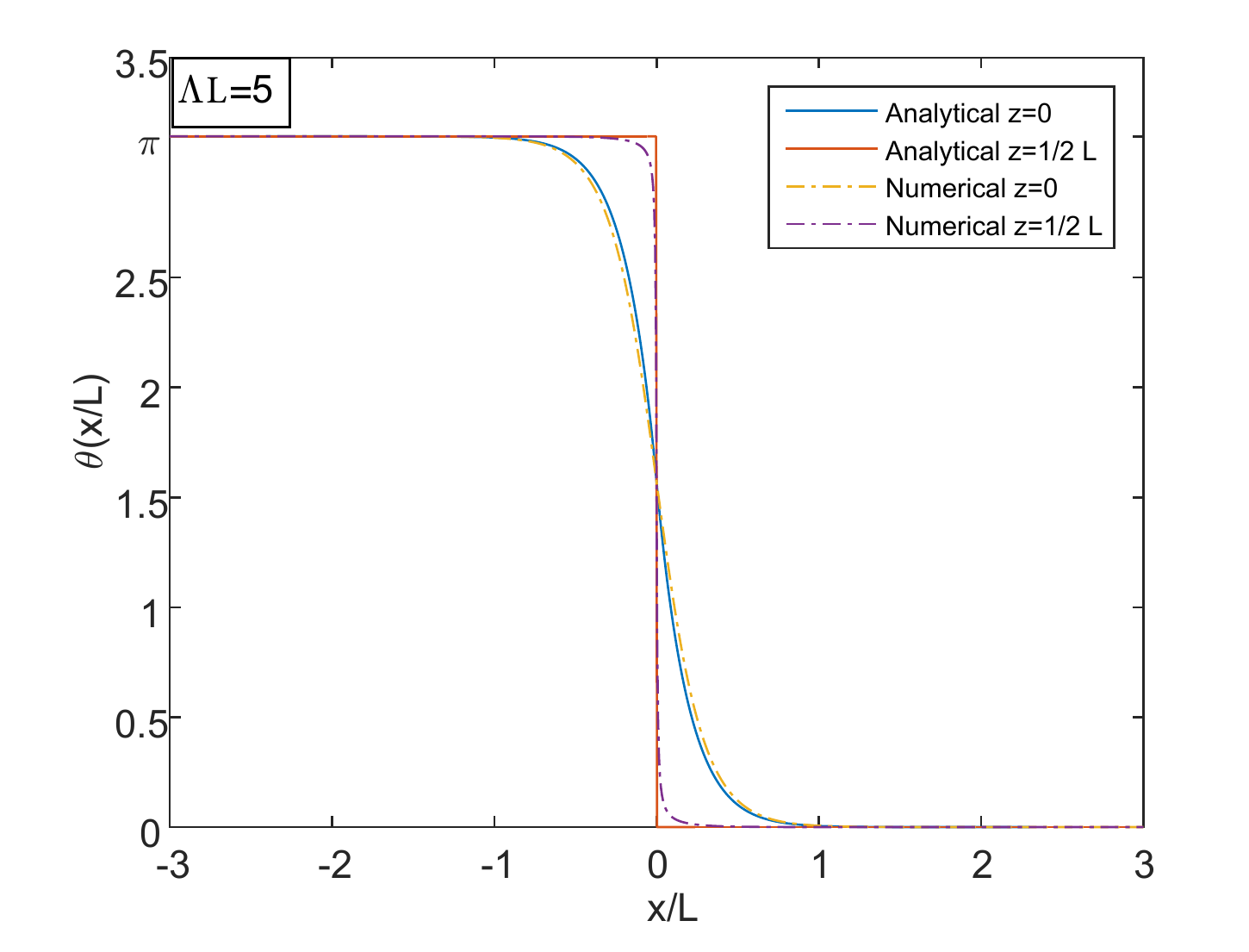}}
\caption{Comparison of the numerical and approximated $\pi$-helicoid solutions for different values of the external field: (a) $\Lambda L = 0$, (b) $\Lambda L = 1$, (c) $\Lambda L = 3$, (d) $\Lambda L = 5$}
\label{profiles}
\end{figure}
\end{center}

\twocolumngrid
\noindent singularities. This alternative approach might be the starting point to study the full nonlinear problem,
  which still presents difficulties to get an explicit formula \cite{FLP} (in particular, the singular behaviour of the solution around the disclination).
  
 In Fig. \ref{piHelicoidsFig},  $\mathbf{n}\lf x, z \rg$ is represented 
   by using formula (\ref{seriessol}) up to $k = 10$. At $\lf 0, \pm \frac{L}{2}\rg$  an overlap of directional  ellipsoids occurs, thus giving rise to  disclinations  extended in the orthogonal direction $y$. 
However, the actual configuration is sensitive to the involved parameters, which control also how good  the linear approximation is. To have more insight in the $\pi$-helicoids,  a numerical boundary value problem solver has been implemented for the full nonlinear system.  
We employed standard numerical techniques, namely the central finite difference scheme accompanied by appropriate application of Newton's iterative method for the calculation of the function $\theta(x,z)$ over a suitable grid \cite{Press2007,LeVeque2007}. The problem can be coded in almost any programming language. However, we used MATLAB$^\copyright$ by Mathworks \cite{matlab} because of the natural way it operates with large and sparse matrices.

In Fig. \ref{profiles} a comparison between the numerical solutions and the analytical series up to 7000 terms is presented. Although the similarity of the profiles is remarkable at this accuracy, a study of the Laplacians shows the expected differences associated to the different equations solved (either modified Helmholtz or sine-Gordon).

%%%%%%%%%%%%%%%%%%%%%%%%%%%%%%%%%%%%%%%%%%%%%%
%%%%%%%%%%%%%%%%%%%%%%%%%%%%%%%%%%%%%%%%%%%%%%
%%%%%%%%%%%%%%%%%%%%%%%%%%%%%%%%%%%%%%%%%%%%%%

\section{Phase transition diagram}

In order to assess the relative stability of the solutions we have found above, {\it i.e.}, $\pi$- and $2\pi$-helicoids, we need now to perform an energy analysis and to compare the solutions also with the uniform nematic configurations to see which  one is energetically favoured in terms of the physical parameters involved, namely $q_0$ and $\Lambda$. To this purpose we use solutions (\ref{sol2pihelic}) for $n=0$ and (\ref{seriessol}) with its extension for negative $x$, where in the latter we truncated the series at $k_{\max} = 7000$.

\begin{figure}[t]
\centering
\includegraphics[width=7cm, height=7cm]{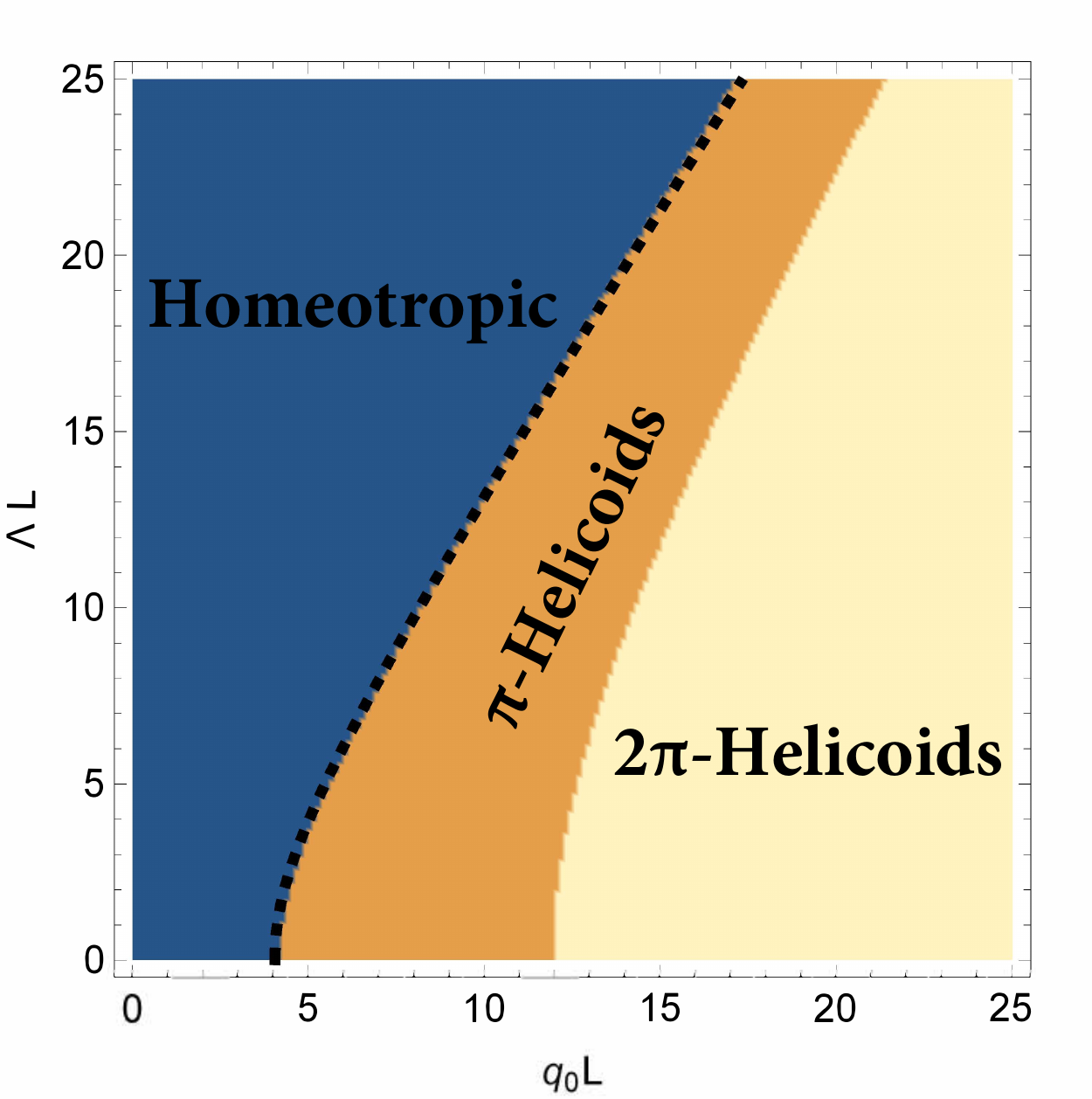}
\caption{Phase diagram in the plane $L q_0, L \Lambda$, representing the nematic (blue), $\pi$-helicoid (orange) and $2\pi$-helicoid (ochre) phases. The dashed black line corresponds to the best fit to \eqref{lambdafit}.}
\label{fig:phasedg}
\end{figure}

Recalling that the chiral strength $q_0$ is a function of the temperature (linear under certain conditions), one could interpret the phase diagram as the result of thermal-magnetic competitive effects in the formation of the helicoids. 
When the energy of both solutions is greater than zero,  the homeotropic nematic phase is the favoured one. It is important to note that the transitions between the three different configurations occur along two different curves (see Fig. \ref{fig:phasedg}). There are two different thresholds in chirality strength, for fixed magnetic field, also depending on the disclination size $a/L$. In Fig. \ref{fig:phasedg}, we have set $a/L=10^{-2}$, although for smaller values of $a/L$ the shape of the diagram does not change.  
In particular, if a study of a pair of $\pi$-helicoids as a function of the distance between them is realised, one can see the energy increasing when they approach. This implies the existence of a barrier preventing the two helicoids from sitting on top of each other, unless some energy is introduced into the system from outside. This underlines the stability of the $2\pi$-helicoids with disclinations located at the boundaries with respect to the decay into two $\pi$-helicoids with disclinations placed at the same point, which correspond to completely different configurations. Conversely, this is not the case when $\Lambda = 0$ where the two configurations are equivalent.

Moreover, the curve between the nematic and $\pi$-helicoid phases should be the analogous of the straight line $\Lambda_0=\frac{\pi | q_0 |}{2} $ in the bulk model \cite{ABC}. Deviations from such a behaviour  are related to the anchoring, possibly leading to significant variations in the value of the critical external field. Moreover, in \cite{Lee92} it was  found a semi-empirical coexistence curve between the homeotropic and the cholesteric phases, providing  the critical external field $\Lambda_c L$ in terms of the cholesteric twist $q_0 L$.  Using the present notation, such a relation reads
 \begin{equation}
\Lambda_c L= \sqrt{
\frac{2 \gamma^2 q L \lf q L-\frac{2}{\gamma} \rg }{ 1-\frac{\lf 1-e^{-\gamma q L}\rg}{\gamma qL}} 
}, 
\label{lambdafit}
\end{equation}
with  $\gamma=K_{2}/K$ and 
\beq
q L = -\frac{1}{\gamma} W_{-1}\left( -\frac{\gamma L}{2 a} e^{-1-q_0 L \gamma}\right),
\label{qeq}
\eeq
where $W_{k}$ is the $k$-th branch of the Lambert $W$ function \cite{DLMF}.
A fitting procedure to our numerical data leads to   $\gamma=1.054$, in excellent agreement with our assumed one constant approximation (\emph{i.e.} $\gamma=1$). The corresponding best fit curve is displayed in Fig. \ref{fig:phasedg} (black dashed line).

%%%%%%%%%%%%%%%%%%%%%%%%%%%%%%%%%%%%%%%%%%%%%%
%%%%%%%%%%%%%%%%%%%%%%%%%%%%%%%%%%%%%%%%%%%%%%
%%%%%%%%%%%%%%%%%%%%%%%%%%%%%%%%%%%%%%%%%%%%%%

\section{ Conclusions}
Summarizing,  we have analytically found   $2 \pi$-helicoids which, to our knowledge, are novel configurations in bounded CLCs with homeotropic anchoring, allowed by the nonlinearity arising from  the presence of an external field. Analogously, we studied   a different type of helicoids, the $\pi$ ones. If the former configuration can be derived in a closed form, for instance by the B\"acklund transformations,  for the latter  we have obtained  approximated expressions and numerical solutions. Both classes of configurations are characterised by the presence of disclinations, located at the boundaries. The disclinations imply energy density divergences, which may be overcome by introducing a  phenomenological energy cut-off, corresponding to excluded regions of melted crystal.

Numerically, we provided a phase diagram in the parameters $q_0 L$ and $\Lambda L$, which established the energetically favoured configurations among them and the uniform nematic phase and the corresponding transitions. We showed that $\pi$-helicoids switch to $2\pi$-helicoids under certain circumstances dictated by the parameters of the problem. In particular, for fixed value of the external field, we found the sequence uniform-to-$\pi$-to-2$\pi$-helicoids as the value of the chirality increases, as expected. 

We would like to specify that other configurations, like the 3-dimensional objects discussed in the Introduction and helicoid lattices, might  modify the phase diagram presented in Figure 5. However, it is not necessary to introduce them in order to characterize the cholesteric-nematic transition we found, as shown by the agreement of our results with those obtained in \cite{Lee92}, and the nature of the disclinations at the boundaries. This approach is similar to previous studies of analytical solutions limited to single helicoids with disclinations but in the absence of external fields and under the assumption of y-invariance (see for instance \cite{goosens, scheffer} and also \cite{17, press1976}). Preceding detailed energy computations are now extended to the case of external fields, in order to draw a theoretical phase diagram which can be considered equivalent to the experimental or semi-empirical ones presented in \cite{Lee92,oswald1,oswald90}.

This work provides a step forward in the analytical construction of the fully comprehensive phase transition diagram for geometrically frustrated chiral nematics, which, to the best of our knowledge, is still a crucial and formidable problem for liquid crystals theory. In this spirit, the next step to be taken would be the study of the existence of lattices of helicoids, their interactions and the corresponding phase diagram.

\bigskip
%%%%%%%%%%%%%%%%%%%%
\section{Acknowledgments}   
The authors are thankful to the Referees for their comments and suggestions. CN is supported by the INFN grant  19292/2017 {\it Integrable Models and Their Applications to Classical and Quantum Problems}.  LM has been partly supported by INFN through the MMNLP project by CSN4. 
%%%%%%%%%%%%%%%%%%%%

\appendix*
\section{}

In this Appendix we give details about the derivation of the integral solution (\ref{integralAFrepr}) to the modified Helmoltz boundary value problem and its relation to the series representation (\ref{seriessol}).

The sine-Gordon Eq. (\ref{SineGordon}) is well known to be solvable via the Inverse Spectral Transform (IST) \cite{AKNS}, for initial data given at a fixed value of one of the independent variables. The IST is an integration procedure, which consists in studying the analytical properties of the wave-functions of a specific linear differential operator (sometimes called the principal Lax operator) in one independent variable.  Then, analytical deformations in the other independent variable of those wave-functions are obtained by the action of a second suitable differential linear operator. In our case, these operators, called Lax pair, possess the key property to have the sine-Gordon as their compatibility (commutativity)  condition. This formulation introduces  a complex spectral parameter $\lambda$, analogous to the momentum/frequency in the Fourier Transform method. In fact,  one may encode information of the solution (like initial data and/or boundary values) into certain specific functions of $\lambda$: the so-called scattering data. The second Lax operator leads to linear  differential equations determining the deformations of the scattering data. This new data is used to reconstruct the solution of the sine-Gordon for any value of the pair of independent variables, by solving the so-called inverse linear spectral problem.  Moreover, the scattering data is determined by the singularities of the principal Lax operator wave-functions in the complex $\lambda$-plane.  Under suitable  asymptotic  assumptions on the sine-Gordon solutions, independent wave-functions  are sectionally holomorphic in the $\lambda$-plane and the corresponding jump functions,    along  the curves  separating the holomorphicity  regions, provide the scattering data. Now, in classical complex function theory, solving the Riemann-Hilbert problem means reconstructing these sectionally holomorphic wave-functions from given jump functions \cite{AblowitzClarkson}. This is  an equivalent way to formulate the inverse spectral problem.  This idea can be readily applied also in the elliptic case, with rather general boundary conditions, provided that suitable jump functions are assured \cite{Unified,AshF}. For the sine-Gordon on the semi-strip  this approach was almost fully developed in \cite{FLP} (and references therein), where a unified method both for linear and nonlinear integrable equations has been developed. Thus, this includes the treatment of the linear (modified) Helmholtz equation, which is exactly the linear approximation to our original model \eqref{SineGordon}. Thus, in the present appendix we will apply such methods to extract interesting information in the linear case given by Eq. (\ref{bvpthetap:1}). This simplification is supported by numerical calculations (see discussion at the end of section \ref{PiHelicoids}).
In fact,  it turns out that linear and nonlinear treatment show quite small differences in the quantitative behaviour of the  solutions near the disclination and for values of $\theta \approx \frac{\pi}{2}$, where we are far from linear scheme.

As mentioned in the main text, in the unified approach to the study of the sine-Gordon and the modified Helmholtz equation, the traditional  Lax pair formulation is equivalent to finding a matrix integration factor $\Psi \lf x, z; \lambda \rg$ (the wave-function),  depending on  $\lambda$, which makes the 1-form $W$  in Eq. (\ref{W_form}) exact.

The first observation is that $W_{lin}$ in Eq. (\ref{W_lin}) remains closed by adding a suitable exact 1-form $ d \lf
  e^{\Omega \, x + \omega\,  z}  \kappa \lf x, z\rg \rg$. In particular, one can choose
the function $ \kappa \lf x, z\rg $ in such a way to cancel the  $\p_z\, \theta$ ($\p_x\, \theta$)  terms from the $dx$  ($dz$) component. A special case is given by \bea W_{lin}^{mod} & =& \frac{ e^{-  \Omega\lf \lambda \rg x -  \omega\lf \lambda \rg z }}{2} \nn \\
& &\lfq     \lf \imath \theta_x -  \Omega\lf \lambda \rg \int_x^{+ \infty }\theta_z\lf \xi, z\rg \, d \xi + \frac{ \Lambda \theta}{\lambda}\rg \right.  \, dx \nn \\ &&  + \left. \lf \imath \theta_z  - \frac{\imath \, \Lambda \theta}{\lambda} +  \int_x^{+ \infty }\theta\lf \xi, z\rg \, d \xi - \right.\right.\nn \\  & & \left. \left.   \omega\lf \lambda \rg \int_x^{+ \infty }\theta_z\lf \xi, z\rg \, d \xi \rg \, dz
  \rgq , \eea
  
\noindent where $\Omega$ and $\omega$ are as in Eq. (\ref{omegas}). 
 
\begin{figure}[thb]\begin{center}\includegraphics[width=0.4\textwidth]{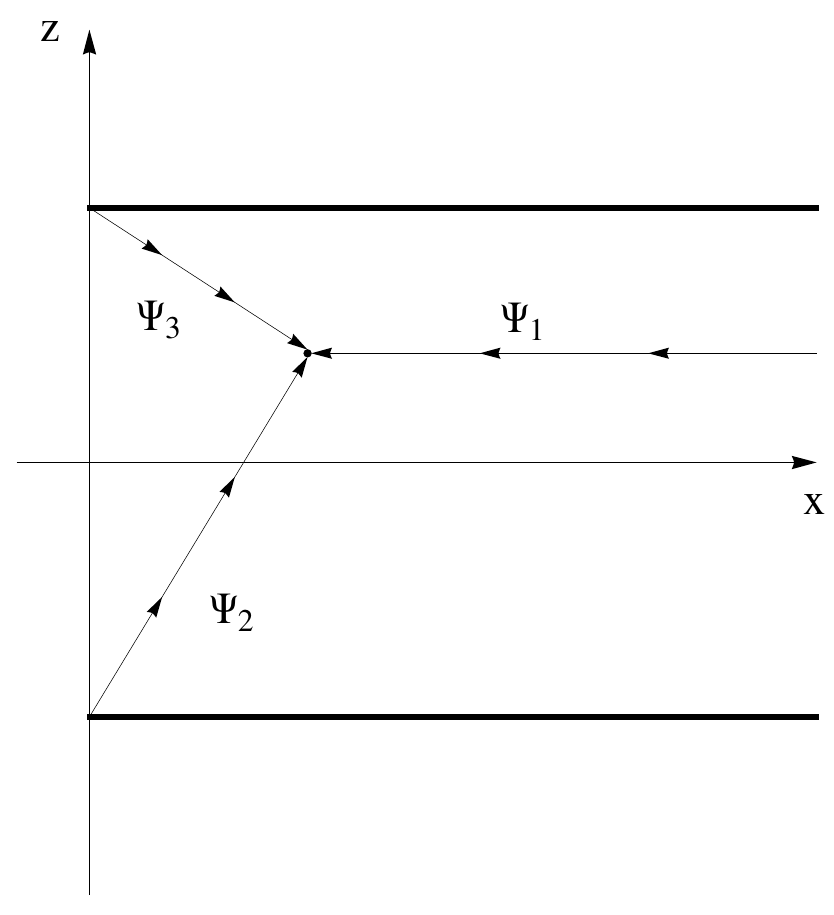}
   \caption{Integration path for the Riemann-Hilbert problem (\ref{c1}) - (\ref{c3}).} \label{integration-pathFig}
\end{center}\end{figure}

Suppose now that the boundary corresponds to a semi-strip where the value for $\theta$ is assigned on its three sides. Then, in analogy with the equation \eqref{W_form},  we look for three functions $\Psi_j, \; j =1, 2, 3$ such that
  \beq d \lf e^{-  \Omega\lf \lambda \rg x -  \omega\lf \lambda \rg z } \Psi_j\rg =  W_{lin}^{mod} .\label{dPsiWlin}\eeq
  By integrating along the paths shown in  Fig \ref{integration-pathFig} (as many as the sides of the semi-strip) with the initial conditions
  \beq \Psi_1\lf +\infty, z; \lambda \rg =  \Psi_{2}\lf 0, - \frac{L}{2}; \lambda \rg=\Psi_{3}\lf 0,  \frac{L}{2}; \lambda \rg = 0 ,\eeq
\noindent for any $\lambda$ one finds the functions $\Psi_j$, which are related by
  \bea \Psi_3 -\Psi_1 &=& -  e^{  \Omega  x  + \omega z } e^{- \omega  \frac{L}{2} } \Psi_1\lf 0,  \frac{L}{2}; \lambda \rg, \label{c1} \\
 \Psi_3 -\Psi_2 &=&  e^{  \Omega  x  + \omega  z } e^{ \omega  \frac{L}{2} } \Psi_3\lf 0,  -\frac{L}{2}; \lambda \rg, \label{c2}\\
 \Psi_1 -\Psi_2 &=& e^{  \Omega  x  + \omega  z } e^{ \omega  \frac{L}{2} } \Psi_1\lf 0,  -\frac{L}{2}; \lambda \rg. \label{c3}
 \eea
 The consistency of these equations requires the global condition
 \beq
 e^{ -\omega  L } \Psi_1\lf 0,  \frac{L}{2}; \lambda \rg - \Psi_1\lf 0, - \frac{L}{2}; \lambda \rg+
 \Psi_3\lf 0,  -\frac{L}{2}; \lambda \rg = 0. \label{globSymm}
 \eeq
If the analyticity properties of the functions $\Psi_j$ are determined and the r.h.s. of \eqref{c1}-\eqref{c3} are known, such a system will define a  Riemann-Hilbert problem associated to the original modified Helmholtz BVP.

Proceeding in this way, one first integrates  equation \eqref{dPsiWlin}  along the path 1, obtaining
 \bea
 \Psi_1 &=& - \half \int_x^\infty\; e^{ \Omega  \lf x -x' \rg} \lfq \imath \theta_{x'}+ \frac{ \Lambda \theta}{\lambda} \right. \nn \\ && \left.  -  \Omega\lf \lambda \rg \int_{x'}^{+ \infty }\theta_z\lf \xi, z\rg \, d \xi \rgq   \, dx'. \label{solPsi1}
 \eea
 Hence the constants appearing in \eqref{c1} and \eqref{c3} are given by
 \bea
 &&\Psi_1\lf 0, \pm \frac{L}{2}; \lambda \rg = \nn \\  &&- \half \int_0^\infty\; e^{ -\Omega  x' } \lfq \imath \theta_{x'}\lf x', \pm \frac{L}{2}\rg  + \frac{ \Lambda \theta\lf x', \pm  \frac{L}{2}\rg}{\lambda} \right. \nn \\ && \left. -  \Omega\lf \lambda \rg \int_{x'}^{+ \infty }\theta_z\lf \xi, \pm \frac{L}{2}\rg \, d \xi \rgq   \, dx' .  \label{Psi10Lm}
\eea

\noindent Since $x' \geq 0$ the above integrals are convergent for $\Im \lambda > 0 $.

Now, in \eqref{c2} the quantity  $\Psi_3\lf 0,  -\frac{L}{2}; \lambda \rg$ can be computed again by integrating \eqref{dPsiWlin}, for $j=3$,  along the segment  $\lfq \lf 0, \frac{L}{2}\rg , \lf 0, -\frac{L}{2}\rg\rgq$. It yields
\bea
 &&\Psi_3\lf 0,  -\frac{L}{2}; \lambda \rg =  \\  &&- \half \int_{-\frac{L}{2}}^{\frac{L}{2}}\; e^{- \omega \lf z'+\frac{L}{2}\rg } \lfq  \imath \theta_{z'}\lf 0, z'\rg  - \frac{\imath \, \Lambda \, \theta\lf 0, z'\rg }{\lambda} + \right.\nn \\   & &  \left. \int_0^{+ \infty }\theta\lf \xi, z'\rg \, d \xi -   \omega\lf \lambda \rg \int_0^{+ \infty }\theta_{z'}\lf \xi, z'\rg \, d \xi  \rgq   \, dz' . \nn
\eea
Integrating by parts the last double integral
\bea   && \half \int_{-\frac{L}{2}}^{\frac{L}{2}}\; e^{- \omega \lf \lambda \rg \lf z'+\frac{L}{2}\rg }\omega\lf \lambda \rg \int_0^{+ \infty }\theta_{z'}\lf \xi, z'\rg \, d \xi     \, dz'  = \nn \\
&& \frac{\omega}{2}\lfq e^{-\omega L} \int_{0}^{\infty}\;\theta\lf \xi,  \frac{L}{2}\rg \, d \xi - \int_{0}^{\infty}\;\theta\lf \xi,  -\frac{L}{2}\rg \, d \xi \rgq \nn \\
&+&\frac{\omega^2}{2}\int_{-\frac{L}{2}}^{\frac{L}{2}}\;e^{-\omega \lf \lambda \rg \lf z'+\frac{L}{2}\rg} \int_{0}^{\infty}\;\theta\lf \xi,  z'\rg \, d \xi\; dz', \eea
the previous expression  reads
 \bea
 &&\Psi_3\lf 0,  -\frac{L}{2}; \lambda \rg =   \half \int_{-\frac{L}{2}}^{\frac{L}{2}}\; e^{- \omega \lf z'+\frac{L}{2}\rg } \times \label{Psi30Lm}\\  && \lfq  -\imath \theta_{z'}\lf 0, z'\rg  + \frac{\imath \, \Lambda \theta\lf 0, z'\rg }{\lambda} - \Omega^2  \int_0^{+ \infty }\theta\lf \xi, z'\rg \, d \xi  \rgq dz'
 \nn \\  & +&    \frac{\omega}{2}\lfq e^{-\omega\, L}\int_0^{+ \infty }\theta\lf \xi, \frac{L}{2}\rg \, d \xi  - \int_0^{+ \infty }\theta\lf \xi, -\frac{L}{2}\rg \, d \xi \rgq   \, dz'
\nn\eea
for any $\lambda \in \mathbb{C}$.

Although the constants introduced in \eqref{c1}-\eqref{c3} contain information about the boundary value problem, they still require unknown data. Hence, further manipulations are needed in order to suppress them.

Turning again our attention to the original problem \eqref{eq:bvpthetap},   Eq. \eqref{Psi10Lm} and  Eq. \eqref{Psi30Lm} become
    \beq \Psi_1\lf 0,  \frac{L}{2}; \lambda \rg =   \frac{\Omega\lf \lambda \rg}{2} \int_0^\infty\; e^{ -\Omega  x' }     \int_{x'}^{+ \infty }\theta_z\lf \xi,  \frac{L}{2}\rg \, d \xi   \, dx'  ,\label{Psi10LmB} \eeq
\bea
\Psi_3\lf 0,  -\frac{L}{2}; \lambda \rg &=&  \frac{i \pi  \left(1 - e^{-\omega\lf \lambda \rg L }\right)}{\lambda ^2+1} \label{Psi30LmB}\\ && - \frac{\Omega^2}{2}  \int_{-\frac{L}{2}}^{\frac{L}{2}}\; e^{- \omega \lf z'+\frac{L}{2}\rg }\int_0^{+ \infty }\theta\lf \xi, z'\rg \, d \xi   dz' ,\nn
\eea

\noindent both still containing  unknown functions.

To have more restrictions  on them, we look at their symmetries. First,  the $z$-mirror  symmetry  \beq \theta\lf x, -z\rg = \theta\lf x, z\rg\eeq
 implies
\beq \Psi_1\lf 0,  \frac{L}{2}; \lambda \rg = - \Psi_1\lf 0, - \frac{L}{2}; \lambda \rg \label{Psi10LmBsymm}\eeq for \eqref{Psi10LmB},
which, used in the global symmetry \eqref{globSymm}, leads to
\beq
 \Psi_3\lf 0,  -\frac{L}{2}; \lambda \rg =  \lf e^{ -\omega  L }  + 1 \rg  \Psi_1\lf 0, - \frac{L}{2}; \lambda \rg. \label{globSymm1}
 \eeq
 Now, since $\Psi_1\lf 0, - \frac{L}{2} ; \lambda\rg$ depends on $\lambda$ only through $\Omega\lf \lambda \rg $, it  will enjoy the same inversion symmetry  $\lambda \to - \frac{1}{\lambda}$, namely
 \beq \Psi_1\lf 0, - \frac{L}{2}; - \frac{1}{\lambda} \rg =  \Psi_1\lf 0, - \frac{L}{2};  {\lambda} \rg.\label{Psi10LmBsymInv}\eeq
On the other hand, $\Psi_3\lf 0, - \frac{L}{2} ; \lambda\rg$ in \eqref{Psi30LmB} depends on $\lambda$ through both $\omega$ and $\Omega^2$, thus being invariant under $\lambda \to  \frac{1}{\lambda}$
\bea \Psi_3\lf 0,  -\frac{L}{2}; \lambda \rg -  \frac{\imath  \pi  \left(1 - e^{-\omega\lf \lambda \rg L }\right)}{\lambda ^2+1} = \nn \\
\Psi_3\lf 0,  -\frac{L}{2}; \frac{1}{\lambda} \rg -
\frac{\imath \,\pi\, \lambda^2 \, \left(1 - e^{-\omega\lf \lambda \rg L }\right)}{\lambda ^2+1} .\label{Psi30LmBsymInv}\eea
Applying the transformation $\lambda \to  \frac{1}{\lambda}$ into \eqref{globSymm1} and \eqref{Psi10LmBsymInv}, substituting into \eqref{Psi30LmBsymInv} and  rearranging the various terms we are led to the equation
\beq
\Psi_1\lf 0, - \frac{L}{2}; {\lambda} \rg -  \Psi_1\lf 0, - \frac{L}{2};  -{\lambda} \rg = G_1\lf \lambda \rg,
\eeq
which, because of the convergence region for $\Psi_1$, only makes sense for $\lambda \in \mathbb{R}$ with
 \beq G_1\lf \lambda \rg = \frac{\imath \pi}{4} \frac{1- \lambda^2}{1+ \lambda^2} \frac{e^{\omega\lf \lambda \rg L} - 1}{e^{\omega\lf \lambda \rg L} + 1}. \label{HR11}\eeq
 Analogously, using again \eqref{globSymm1} and the above symmetries, one obtains
 \beq
\Psi_3\lf 0,  -\frac{L}{2}; \lambda \rg -  \lf e^{ -\omega  L }  + 1 \rg  \Psi_1\lf 0, - \frac{L}{2}; -\lambda \rg =  G_2\lf \lambda \rg, \label{HR31}
 \eeq
which holds on $\lambda \in \mathbb{C}^-$ and where
\beq \;G_2\lf \lambda \rg =  \frac{\imath \pi}{4} \frac{1- \lambda^2}{1+ \lambda^2} \lf 1 - e^{- \omega\lf \lambda \rg L} \rg .  \label{HR22}
\eeq
 Both $G_1$ and $G_2$ now only encode information about the  boundary conditions, but in order to use them one has to suitably modify   the relations \eqref{c1}-\eqref{c3}. Precisely, defining the new $\tilde{\Psi}_j$ (coherently with the definition \eqref{dPsiWlin})
\beq
\begin{array}{lc}
  \Psi_1 = \tilde{\Psi}_1&    \lambda \in \mathbb{C}^+  ,   \\
\Psi_2 = \tilde{\Psi}_2 - e^{\Omega\, x + \omega\, \lf z + \frac{L}{2}\rg}  \Psi_1\lf 0, - \frac{L}{2}; -\lambda \rg      &\lambda \in \mathbb{C}^{III}   ,  \\
   \Psi_3 = \tilde{\Psi}_3 + e^{\Omega\, x + \omega\, \lf z-\frac{L}{2}\rg}   \Psi_1\lf 0, - \frac{L}{2}; -\lambda \rg &      \lambda \in \mathbb{C}^{IV},
\end{array}
\eeq
the  relations \eqref{c1}-\eqref{c3} read
 \beq
\begin{array}{lccr}
\tilde{\Psi}_3 -\tilde{\Psi}_1 &=& e^{  \Omega  x  + \omega z } e^{- \omega  \frac{L}{2} } G_1\lf \lambda\rg , &\lambda \in \mathbb{R}^+ , \\
 \tilde{\Psi}_3 -\tilde{\Psi}_2 &=&  e^{  \Omega  x  + \omega  z } e^{ \omega  \frac{L}{2} } G_2\lf \lambda \rg, &\lambda \in \imath \mathbb{R}^-  , \\
\tilde{\Psi}_1 -\tilde{\Psi}_2 &=& e^{  \Omega  x  + \omega  z } e^{ \omega  \frac{L}{2} } G_1\lf \lambda\rg , & \lambda \in \mathbb{R}^- .
\end{array} \label{RHp}
\eeq
  The system \eqref{RHp} is a Riemann-Hilbert problem defined on three branches, where the jumps are completely known functions.
   Furthermore,  going back to \eqref{solPsi1} and estimating its asymptotic behaviour for $\lambda \to \infty$, one obtains a reference value for $\tilde{\Psi}_1$
  \beq
  \tilde{\Psi}_1\lf x, z; \lambda \rg = \half \int_x^{+ \infty }\theta_z\lf \xi, z\rg \, d \xi +O\lf \frac{1}{\lambda}\rg.\label{asyPsi1}
  \eeq
 Thus, from the jump conditions  \eqref{RHp} and the asymptotic value \eqref{asyPsi1}, $\tilde{\Psi}_1$  is given by
 \bea
 &&2 \pi \imath\ \lfq \tilde{\Psi}_1\lf x, z; \lambda \rg - \half \int_x^{+ \infty }\theta_z\lf \xi, z\rg \, d \xi \rgq  =  \nn \\
&& \int_{-\infty}^0 e^{  \Omega\lf \lambda'\rg   x  + \omega\lf \lambda'\rg \lf   z+ \frac{L}{2}\rg } \frac{G_1\lf \lambda'\rg }{\lambda' - \lambda} d\lambda' \nn \\
&& + \int_{\infty}^0 e^{  \Omega\lf \lambda'\rg   x  + \omega\lf \lambda'\rg \lf   z- \frac{L}{2}\rg } \frac{G_1\lf \lambda'\rg }{\lambda' - \lambda} d\lambda' \nn \\
&& + \int^{-\imath \infty}_0 e^{  \Omega\lf \lambda'\rg   x  + \omega\lf \lambda'\rg \lf   z+ \frac{L}{2}\rg } \frac{G_2\lf \lambda'\rg }{\lambda' - \lambda} d\lambda' ,
 \eea
which holds for $\lambda \in \mathbb{C}^+$.
On the other hand, from the component $dx$ of the linear  problem \eqref{dPsiWlin}, or from the solution \eqref{solPsi1},    in the limit $\lambda \to 0$ one obtains
\beq
 \theta = - \imath \lim_{\lambda \to 0}  \lfq \tilde{\Psi}_1\lf x, z; \lambda \rg - \half \int_x^{+ \infty }\theta_z\lf \xi, z\rg \, d \xi \rgq.
\eeq
Comparing the last two relations, one obtains the solution of the modified Helmholtz BVP \eqref{eq:bvpthetap}
 by the  final  formula
\begin{widetext}
\bea \label{appendixSol} \theta_+(x,z)&=& \frac{-1}{2 \pi} \lq  \int_{-\infty}^0 e^{\Omega\lf \lambda \rg x + \omega\lf \lambda \rg \lf z+\frac{L}{2} \rg} G_1\lf \lambda \rg \frac{d\lambda}{\lambda} + \int_{\infty}^0 e^{\Omega\lf \lambda \rg x + \omega\lf \lambda \rg \lf z-\frac{L}{2} \rg} G_1\lf \lambda \rg \frac{d\lambda}{\lambda}\right.\nn \\ && \left. +
\int^{- \imath \infty}_0 e^{\Omega\lf \lambda \rg x + \omega\lf \lambda \rg \lf z+\frac{L}{2} \rg} G_2\lf \lambda \rg \frac{d\lambda}{\lambda}\rgq \label{integralAFreprApp}.\eea
\end{widetext}
 
\noindent One can check that the solution \eqref{appendixSol} certainly satisfies the boundary conditions in \eqref{eq:bvpthetap}. 

Now, if one sets $z = - \frac{L}{2} + \epsilon$ with $0 < \epsilon < \frac{L}{2}$, it can be verified that the integrands in \eqref{integralAFrepr} are bounded and analytic functions  in  $C^{IV}$ .  Thus,  performing as above, the change of variable $\lambda \to - \frac{1}{\lambda}$ on the first term, we are led to an integrand split into an analytic part plus a meromorphic contribution $ - \imath \frac{e^{\Omega\lf \lambda\rg x}}{2 \lambda} \frac{ \left(1 - \lambda ^2\right) }{  \left(1 + \lambda ^2\right) } \frac{\left(1- e^{- \omega\lf \lambda\rg \epsilon}\right)}{\left(e^{L
   \omega\lf \lambda\rg}+1\right)}$, containing all poles in the family $P_G^- = \lgr \lambda_ \in P_G , \; \Im \lambda_n < 0 \rgr$.  Thus, the solution is given in terms of the series of its residues.   Noting  that
\beq \lambda_{-\lf n+1\rg} = - \lambda_{n} - 2 \imath \sqrt{1 + \lf 2 n+1\rg^2 \pi^2} , \eeq
\beq  \Omega\lf \lambda_{-\lf n+1\rg}\rg = \Omega\lf \lambda_{n}\rg, \nn \eeq 
\beq \omega\lf \lambda_{-\lf n+1\rg}\rg = - \omega\lf \lambda_{n}\rg, \nn \eeq 
\beq \textrm{Res}\left(e^{L \omega\lf \lambda\rg}+1\right)^{-1} |_{\lambda_{-\lf n+1\rg}} = - \textrm{Res}\left(e^{L \omega\lf \lambda\rg}+1\right)^{-1} |_{\lambda_{n}} -2 ,\nn \eeq
one can first sum up the contributions coming from the poles $\lambda_{n}$ and $\lambda_{-\lf n+1\rg}$, collecting the exponential $x$-dependence and the trigonometric $z$-dependence. This manipulation leads directly to the formula \eqref{seriessol}.

%%%%%%%%%%%%%%%%%%%%%%%%%%%%%%%%%%%

  \end{document}